\documentclass{aa}
\usepackage{graphicx,amssymb}
\usepackage{natbib}

\begin{document}

\title{Modeling CHANDRA Low Energy Transmission Grating Spectrometer
       Observations of Classical Novae with PHOENIX \\
       I. V4743 Sagittarii}
\titlerunning{Modeling the CHANDRA LETGS Observations of Classical
              Novae with PHOENIX}

\author{A. Petz \inst{1}, P. H. Hauschildt \inst{1}, J.--U. Ness
        \inst{1} \and S. Starrfield \inst{2}}
\authorrunning{Petz et al.}

\institute{Hamburger Sternwarte,
       Gojenbergsweg 112, 21029 Hamburg, Germany\\
       Email: {\tt [apetz;yeti;jness]@hs.uni-hamburg.de}
       \and
       Department of Physics and Astronomy,
       Arizona State University, Tempe, AZ 58287--1504, USA\\
       Email: {\tt starrfield@asu.edu}}

\date{Received date / Accepted date}

\abstract{
We use the {\tt PHOENIX} code package to model the X--ray spectrum of
Nova V4743 Sagittarii observed with the LETGS onboard the Chandra
satellite on March 2003. Our atmosphere models are 1D spherical,
expanding, line blanketed, and in full NLTE. To analyze nova
atmospheres and related systems with an underlying nuclear burning
envelope at X--ray wavelengths, it was necessary to update the code
with new microphysics, as discussed in this paper. We demonstrate that
the X--ray emission is dominated by thermal bremsstrahlung and that
the hard X--rays are dominated by Fe and N absorption. The best fit to
the observation is provided at a temperature of
T$_{\rm eff}$ = $5.8 \times 10^5$\,K, with
L$_{\rm bol}$ = 50 000 L$_\odot$. The models are calculated for solar
abundances. It is shown that the models can be used to determine
abundances in the nova ejecta.}

\maketitle


\section{Introduction}

Classical novae (CNe) are the third most violent of the stellar
explosions that can occur in a galaxy (after $\gamma$--ray bursts and
supernovae). About 5 to 10 Galactic CNe are discovered per year in the
Solar neighborhood. CNe participate in the cycle of Galactic chemical
evolution, in which they eject grains and metal--enriched gas as a
source of heavy elements for the interstellar medium
\cite[supplementing those of supernovae, AGB stars, PNe, and WR
stars:][]{gehrz98}. In addition, CNe are related to the Super Soft
Binary X--ray Sources (SSS) which are probably progenitors of SN Ia
explosions \cite[]{kahabka97} and which are thought to be the result
of thermonuclear explosions in the cores of mass--accreting white
dwarfs in close binary systems
\cite[]{hillebrandt00,leibundgut01}. The association of CNe and SSS is
important both because of the cosmological implications of SN Ia
(currently the best standard candles for determining cosmological
distances) and because SN Ia are thought to be responsible for the
abundance of iron in the solar system.\par
CN explosions occur in a  cataclysmic variable (CV) binary star system
in which a Roche lobe filling secondary supplies hydrogen--rich
material that is accreted onto a white dwarf (WD) primary. Theoretical
studies show that the accreted layer grows until it is sufficiently
hot and dense to initiate a thermonuclear runaway (TNR). The evolution
of the TNR depends upon the mass and luminosity of the WD, the rate of
mass accretion, the  composition of the accreted material, and the
chemical composition in the reacting layers
\cite[]{starrfield89,gehrz98}. Observations imply and theory demands
that core material be dredged--up into the accreted material and the
gases be thoroughly mixed before being ejected into space
\cite[]{starrfield98,gehrz98}. {\it Thus, the chemical composition of
the ejected material reflects a TNR processed mixture of WD core plus
accreted material.}  Abundance studies of the ejecta carried out with
the IUE satellite have established that the underlying WDs are either
carbon--oxygen (CO) or oxygen, neon, and magnesium (ONeMg) WDs
\cite[]{starrfield98,gehrz98}. CNe are expected to be the major source
of $^{15}$N and $^{17}$O in the Galaxy and to contribute to the
abundances of other isotopes in this atomic mass range.\par
Previous X--ray studies of CNe in outburst show {\it that no other
wavelength region provides unambiguous information on the evolution
and characteristics of the underlying WD}
\cite[]{drake03,krautter96,krautter02,ness03}. For example, the ROSAT
studies discovered a ``soft'' component in CNe and placed them at, or
near, the bright end of the class of SSS. The ``soft'' component has
been identified as emission from the hot nuclear burning photosphere
of the WD and theoretical studies show that a determination of the
temperature and lifetime is a measure of the mass of the WD. Stellar
atmosphere analyses of the photosphere can provide abundances of the
material remaining on the WD. CNe at optical maximum are luminous,
L$\ge$ $10^5$L$_\odot$ and evolve to X--ray maximum where T$_{\rm
eff}$ ranges from $3\times 10^5$K to 6$\times 10^5$K or higher
\cite[]{krautter96,balman98,orio02,orio03}.\par
X--ray emission during the CN outburst can be divided into three
phases \cite[]{krautter02b}. First is the early fireball phase, in
which the hot atmosphere of the white dwarf is expanding and cooling
adiabatically.  The peak temperature of the white dwarf is predicted
to exceed $10^6$K \cite[depending on the mass of the white
dwarf:][]{starrfield96} so, if detected in this phase, it would be an
extremely hot source. Nevertheless, the expanding shell cools rapidly
and becomes more opaque in X--rays in a few hours. Therefore, for the
first few hours of the outburst, after the shell has cooled
sufficiently, the nova is probably undetectable by an X--ray
satellite.  However, after a few days, the expanding shell starts to
become ionized and the copious X--rays from the inner layers partially
penetrate the shell and X--ray observations of novae during this phase
show emission lines from the expanding gas
\cite[]{mukai01,orio02,orio03}. This is the second phase.\par
The third is the constant bolometric luminosity phase when the system
again becomes observable as a ``soft'' source. This phase occurs
because both observational and theoretical studies of CNe show that
not the complete envelope is ejected during the explosive phase of the
outburst. Some fraction remains on the WD and rapidly returns to
hydrostatic equilibrium. This material provides sufficient fuel so
that the enlarged WD remains hot and luminous for months to years
\cite[]{krautter96}. It is predicted that the duration of this phase
is an inverse function of the mass of the WD \cite[]{krautter96} so
that a determination of how long a nova is ``on'' in X--rays can
provide an estimate of the WD mass. Moreover, since the burning WD is
hot, with nuclear burning ongoing in its surface layers, its
characteristics closely resemble those of the SSS.\par
In this paper we report on the analysis of an observation of nova
V4743 Sgr with the High Resolution Camera and the Low Energy
Transmission Grating Spectrograph (HRC--S + LETGS) on the CHANDRA
X--ray satellite \cite[]{ness03}. Because the structure of the
rekindled WD in V4743 Sgr consists of a nuclear burning envelope with
an very hot stellar atmosphere on top, we have updated the
microphysics and numerics in {\tt PHOENIX} to analyze nova atmospheres
at X--ray wavelengths. We describe these improvements in section
\ref{addphys}. {\tt PHOENIX} is well suited to analyze V4743 Sgr, or
any SSS, because it has been developed specifically to treat radiation
transport in an expanding, optically thick atmosphere at any
temperature. We show that we can now treat the formation and time
evolution of the X--ray radiation in a CN in outburst and that it is
possible to determine the chemical composition of the CN ejecta. We
describe the X--ray observations of V4743 Sgr in section \ref{obs},
and end with a Summary and Discussion (sections \ref{res} and
\ref{concl}).

\section{Modeling novae in X--rays}

The nova models calculated with the {\tt PHOENIX} code are 1D
spherical, expanding, line blanketed, NLTE model atmospheres. The
spherical, co--moving frame radiative transfer equation (SSRT) for
expanding media is solved for lines and continua, coupled with the
NLTE statistical equilibrium equations \cite[]{hauschildt99}. The
numerical solution of the radiation transport and multi--level NLTE
problems are based on an operator splitting (ALI) method as described
in detail by \cite{hauschildt92} and \cite{hauschildt99}. Nova
atmospheres are far from LTE, therefore, all models have to include
NLTE for as many atoms and energy levels as feasible.\par
The radiative transfer problem is coupled to the energy conservation
through the equation of radiative equilibrium (in the co--moving
frame) and, therefore, a temperature correction procedure
\cite[]{hauschildt03} is used to iteratively correct the structure so
that energy is conserved.  Mechanical energy sources, due to the
expansion of the nova envelope, as well as convection are negligible
for the conditions found in nova atmospheres.\par
The density profile and the velocity field of the expanding medium are
taken from hydrodynamic simulations \cite[]{shara89,starrfield89}. In
this case the radiation transport problem effectively decouples from
the hydrodynamic equations and the problem is dramatically
simplified. Earlier nova models in other wavelength ranges had shown
that there are satisfactory results by the use of these
simulations \cite[]{hauschildt95b}. Therefore we use the results of
the simulations as a first assumption in this stage of our work.\par
The nova atmosphere is approximated by an expanding but stationary (in
time) structure. This implies that the explicit time dependencies in
the radiation transfer and hydrodynamics (or in the given density
profile and velocity field) can be neglected and that the time
evolution of the nova atmosphere can be represented by a sequence of
snapshot models. This assumption is valid since the hydrodynamic time
scale is much greater than both the radiative time scale and also the
excitation and ionization equilibrium time scales \cite[]{bath76}.\par
To account for LTE and NLTE line blanketing, the opacities of all
important spectral lines have to be included. This also includes the
handling of line blends, e.g., due to line broadening. Doppler
broadening due to the large scale velocity field is handled through
the co--moving frame radiation transport and the Lorentz
transformation of the radiation field is used to obtain the observed
spectrum in the Euler frame. {\tt PHOENIX} includes a large number of
NLTE lines as well as LTE background lines of species not treated in
NLTE \cite[]{hauschildt93,hauschildt95}. It solves the line radiative
transfer equations without using approximations such as the Sobolev
method \cite[see][and ref. therein]{grinin01}. Therefore, all
depth--dependent spectral line profiles have to be calculated in the
Lagrangian frame. This in turn requires a large number of wavelength
points (400,000 for the models presented in this paper).\par
The generation of a model is very time consuming due to the complexity
of the problem. The construction of detailed models is possible only
through the use of parallel algorithms in the {\tt PHOENIX} code
\cite[]{hauschildt97,baron98,hauschildt01}. Therefore, modern parallel
computers are used to allow more complex, more sophisticated, and more
realistic models.\par

\subsection{Physical parameters of a nova model atmosphere}

There are several physical parameters of a nova model
atmosphere. First there is the effective temperature
$T_{\rm eff}$. Together with the reference radius $r_{\rm ref}$
(radius at $\tau_{std} = 1$, where the optical depth grid is
calculated at a standard wavelength, depending on the effective
temperature of the model) it defines the bolometric luminosity of
the nova via $L_{\rm bol} = 4 \pi \sigma T_{\rm eff}^4 r_{\rm
ref}^2$. This definition is only used as a convenient
parameterization.\par
There is ample evidence from modeling of the outburst that the nova
wind is driven by radiation pressure \cite[]{hauschildt94}. Therefore,
the static Eddington luminosity $L_{\rm edd}$ cannot be used as an
upper limit. Thus the luminosity of the nova has to be considered as a
free parameter or taken from hydrodynamical simulations.  For the
models presented here, we have assumed a bolometric luminosity of
$L_{\rm bol} = 50,000 L_{\odot}$. This value implies a white dwarf
mass close to the Chandrasekhar limit. In general, nova model
atmospheres for earlier phases of the outburst have shown very little
dependence of the spectrum on the bolometric luminosity
\cite[]{hauschildt95b}. This is also true for the models presented in
this work and we parameterize our models by the effective
temperature.\par
The density profile of the nova atmosphere can be parameterized as a
power law of the form
\begin{equation}
  \rho(r) \propto r^{-N} .
  \label{densityprofile}
\end{equation}
The velocity profile follows from the requirement that the mass loss
rate $\dot M(r,t)$ is constant in space \cite[]{bath76}, i.e.,
\begin{equation}
  v(r) = \frac{\dot{M}}{4 \pi r^2 \rho(r)}
  \label{velfield}
\end{equation}
with two parameters $N$ and $v_{\rm out}$. For novae, we use a typical
value of $N = 3$. $v_{\rm out}$ is the outer velocity of the expanding
shell, here we use the observed value of
$v_{\rm out} = 2500$\,km s$^{-1}$ determined from the width of the
line features of the X--ray spectra shown below. $\rho_{\rm out}$
determines the outer edge of the envelope and is given by an
outer pressure $p_{\rm out}$ assuming that the ideal gas equation is
valid. Reasonable values lie in the range between
$p_{\rm out} = 10^{-2}$ and $10^{-1}$\,dyn cm$^{-2}$. Together with
$N = 3$ and $r^2$ in Eq.~\ref{velfield} this leads to a velocity field
which is linear in $r$\par

\subsection{Additional physics included in {\tt PHOENIX} for modeling
            novae in the SSS phase \label{addphys}}

In order to model the X--ray spectra of novae with {\tt PHOENIX}, we
had to extend the opacity and line databases into the X--ray spectral
range (from $\lambda \approx 1$\,{\AA}, equivalent to
$E \approx 12.4$\,keV). Several new atomic data and physical processes
have to be included in order to construct detailed nova atmosphere
models with effective temperatures in excess of $10^5\,$K. Previously
the only database with X--ray data available in {\tt PHOENIX} was
CHIANTI Version 3 \cite[]{dere97,dere01}. For this work, we have
extended {\tt PHOENIX} to also include the CHIANTI Version 4
\cite[CHIANTI4,][]{young03} and the APED (Astrophysical Plasma
Emission Database, Version 1.3.1)
\footnote{\tt http://cxc.harvard.edu/atomdb/ \rm} databases.\par
The new databases provide the following extensions compared to
previous {\tt PHOENIX} versions:
\begin{itemize}
  \item NLTE model atom data for highly ionized species or species
with large ionization potential, including
  \item many new spectral lines in the X--ray waveband down to 1 {\AA}
  \item improved data for electron collision rates
  \item new data for proton collision rates
  \item better data for thermal bremsstrahlung
\end{itemize}
All these improvements were used for the models presented in this
work.\par

\begin{table}[t]
  \begin{center}
    \begin{tabular}{|l|l|l|}\hline
      database        & X--ray lines      & all lines         \\ \hline
      CHIANTI Ver.~3  & $\approx 12.650$  & $\approx 45.000$  \\ \hline
      CHIANTI Ver.~4  & $\approx 22.350$  & $\approx 74.500$  \\ \hline
      APED (total)    & $\approx 326.000$ & $\approx 463.200$ \\ \hline
      APED (distinct) & $\approx 178.000$ & $\approx 310.400$ \\ \hline
    \end{tabular}
  \end{center}
    \caption[]{\label{xray_lines} Number of spectral lines from the
    atomic databases. Given are the number of X--ray lines (defined as
    lines with wavelengths $\lambda \le 100$\,{\AA}) and the sum of
    all lines over the whole spectral range (up to
    $\lambda \approx 10^7$\,{\AA}). In the APED database entries for
    the same transition are possible. In the last row the number of
    unique lines are specified.}
\end{table}

In Table \ref{xray_lines} we give the number of atomic lines found in
each of the databases.  The APED database provides a good extension to
the other databases. Note, that  many of the lines in the databases
are rather weak but, collectively, are important opacity
sources. There are a number of transitions with multiple entries for
the same line but different atomic data and it is not clear which
entries are better than others. For this work, we simply use the first
entry for each unique transition of the line databases and ignore all
duplicate data. The effects of using different data where available
will be investigated in a later paper.\par
The values of the electron and proton collision rates for each ion
depend significantly on the database. In addition, the proton
collision rates are only available in the CHIANTI4 and APED
databases. All applicable collision rates are used in the rate
equations of {\tt PHOENIX} for ions whose transitions are treated in
NLTE. A detailed investigation of the sensitivity of the results on
the collisional data will be presented in a subsequent paper.\par
Thermal bremsstrahlung is calculated using the data for the ff--gaunt
factors available in the databases. Previously, {\tt PHOENIX}
considered only contributions of H, Si, and Mg as bremsstrahlung
processes (called 'old bremsstrahlung' in this paper). In the CHIANTI
Version 4 database, the ff--gaunt factors are available for elements
up to Z = 30. These bremsstrahlung processes have been included in
this work (called 'new bremsstrahlung' in this paper) which is
important for highly ionized abundant elements.  The nova model
atmospheres are optically thick, therefore, free--free absorption as
the inverse process to the bremsstrahlung emission is important and
included in the models.  Two ranges for the plasma temperature $T$ and
the ratio of photon to thermal energy $u$ have to be
distinguished. For $T \ge 10^6$\,K and
$u = \frac{h \nu}{kT} \ge 10^{-4}$ (where $\nu$ is the frequency of
the radiation) the relativistic ff--gaunt factors and the fits of
\cite{itoh00} are used. Outside this range we use the ff--gaunt
factors of \cite{sutherland98} since they are valid over a large range
of temperatures and frequencies. In both references, the ff--gaunt
factors are given for elements up to Z = 28. However, it is possible
to enhance the fits up to Z = 30. These extensions are provided in the
CHIANTI4 database and used in the model calculations presented
here.\par

\subsection{Computation of hot nova models}

The models computed for this work are the first {\tt PHOENIX} nova
atmosphere models with effective temperatures
$T_{\rm eff} \ge 10^5$\,K. Due to the extreme conditions in nova
atmospheres, it is better to start the model calculations from scratch
rather than using scaled structures from our earlier nova modeling.
The first step in this process is to calculate an LTE continuum model
without any spectral lines for each temperature, typically starting
from a grey temperature structure. In the next step,  spectral lines
are included in the LTE model to generate a model structure that
includes LTE line blanketing. Next, NLTE models are constructed with
more and more species treated in NLTE. It is important to treat the
transitions of all important ionization stages of one element in NLTE
simultaneously, because NLTE effects like over- and under--ionization
affect the level populations of adjacent ionization stages. Therefore,
large sets of NLTE species are needed in the model
atmospheres. However, there are convergence problems if the
calculation includes too many 'fresh' NLTE species at once. Therefore,
we use a multi--stage process to reach the final models with the full
set of NLTE species.\par
Each step needs about 30 iterations and we calculate about 270
iterations in total for a model presented in this publication. The
model includes about $7\times10^3$ b--f and about $100\times10^3$ b--b
transitions in NLTE. The atmosphere is divided into 50 layers
and distributed along an optical depth grid from
$\tau_{std\_min} = 10^{-10}$ to
$\tau_{std\_max} = 10^{2}$.

\section{X--ray observations of nova V4743 Sgr \label{obs}}

Nova V4743 Sgr was first detected on 20 September 2002 by
\cite{haseda02} at m${_V} = 5$. An accurate position is RA(J2000) =
$19^h01^m09^s.38$, DEC(J2000) = $-22^{\circ}00'05''.9$
\cite[determined by][]{tanaka02}. From the optical light curve the
nova was classified as very fast. A velocity of 2400\,km s$^{-1}$ was
estimated from the full width at half maximum (FWHM) of the H$\alpha$
emission line, indicating that the expansion velocities are
$\sim$ 1200\,km s$^{-1}$. \cite{lyke02} determined a distance of
6.3\,kpc based on infrared observations. The first X--ray observation
of V4743 Sgr was obtained with the ACIS--S instrument on CHANDRA in
2002 November. It was faint at this time and did not appear to be in
the SSS phase as yet.  We then observed it for 24.7 ks with the
HRC--S+LETG on 19 March 2003
\cite[OBSID 3775,][]{ness03,starrfield03}.\par
The light curve for the first grating observation proved extremely
interesting with both large amplitude oscillations and, about
two--thirds of the way through the observation, a rapid decline to
almost no counts per second \cite[]{ness03}. Two further HRC--S+LETGS
observations were carried out on 2003 July 18, and 2003 September 26
(11.7\,ks: OBSID 3776 and 12.0 ks: OBSID 4435).\par

\begin{figure}
  \center
  \includegraphics[width=8cm]{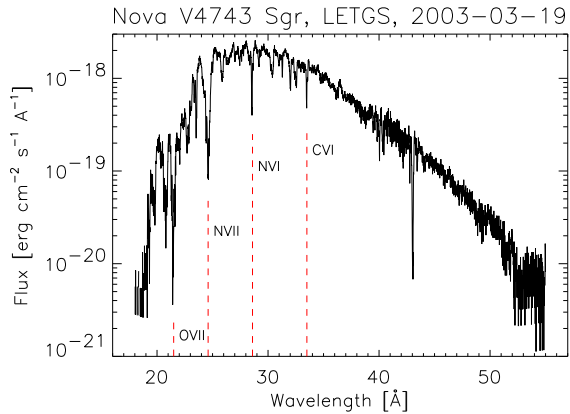}
  \caption[]{\label{observed_spectrum} The observed Chandra LETGS
  spectrum of nova V4743 Sgr. There are absorption lines and probably
  weak emission lines in the spectrum. The strongest lines are from
  the two highest ionisation stages of C, N, and O
  \cite[]{ness03}.}
\end{figure}

The spectrum from the first observation is shown in
Fig.~\ref{observed_spectrum}. The effective areas used to convert from
ct s$^{-1}$ to flux are determined with the CIAO (Chandra Interactive
Analysis of Observations) software
package\footnote{\tt http://cxc.harvard.edu/ciao/ \rm}, version
3.0. The spectrum is background subtracted. At wavelengths greater
than $\approx 55$\,{\AA}, the spectrum is dominated by second and
higher dispersion orders \cite[]{ness03}, and these wavelengths will
not be considered in our analysis.\par
An examination of the spectrum shows that it is not a black--body but
resembles a stellar atmosphere with deep absorption features and,
possibly, some weak emission lines. This spectrum only slightly
resembles that of either CAL 83 \cite[]{paerels01} or V1494 Aql
\cite[]{starrfield01}. The strongest lines are from the two highest
ionisation stages of C, N, and O \cite[]{ness03}.\par
An extensive analysis of the observation from March has been carried
out by \cite{ness03}. A remarkable feature in the light curve from
that observation is a strong decline of the X--ray emission between 13
and 20 ks. Fig.~1 of \cite{ness03} shows the light curve taken during
this observation. It is oscillating at high amplitude with a period of
$\sim$ 22 min. Moreover, the oscillations exhibit spectral variability
and a correlated change in hardness ratio, defined by $(h-s)/(h+s)$,
with $h$ being the flux in 19 \AA{} $< \lambda <$ 30 \AA{} and $s$
being the flux in 30 \AA{} $< \lambda <$ 50 \AA. There is a strong
period present in the data, with at least two harmonic overtones,
which reflect the complex nature of the variations. This is in line
with the X--ray oscillations of V1494 Aql which were also
multiperiodic and were interpreted as pulsation by \cite{drake03}. If
the oscillations in V4743 Sgr (which exhibited a much larger amplitude
than those in V1494 Aql) are interpreted as a rotational modulation
then they imply a major asymmetry in the X--ray emitting region.\par
The spectrum of the last 5.4 ks can be seen in Fig.~4 of
\cite{ness03}. There are strong emission lines but there is no
continuum present so we cannot discern  if absorption lines are
present. In any case, the peak count rate at this time is far lower
than in any of the X--ray observations where a strong continuum is
present. A possible scenario for the decline at 13 ks involves an
eclipse of the X--ray emitting WD. But this does not explain either
the long duration of the decline or the softening during the
decline. The emission spectrum at minimum is probably produced by the
ejected shell and not a hot stellar photosphere. Therefore, we do not
consider it any further in this paper.\par

\section{Results \label{res}}

\begin{figure}
  \includegraphics[width=8.4cm]{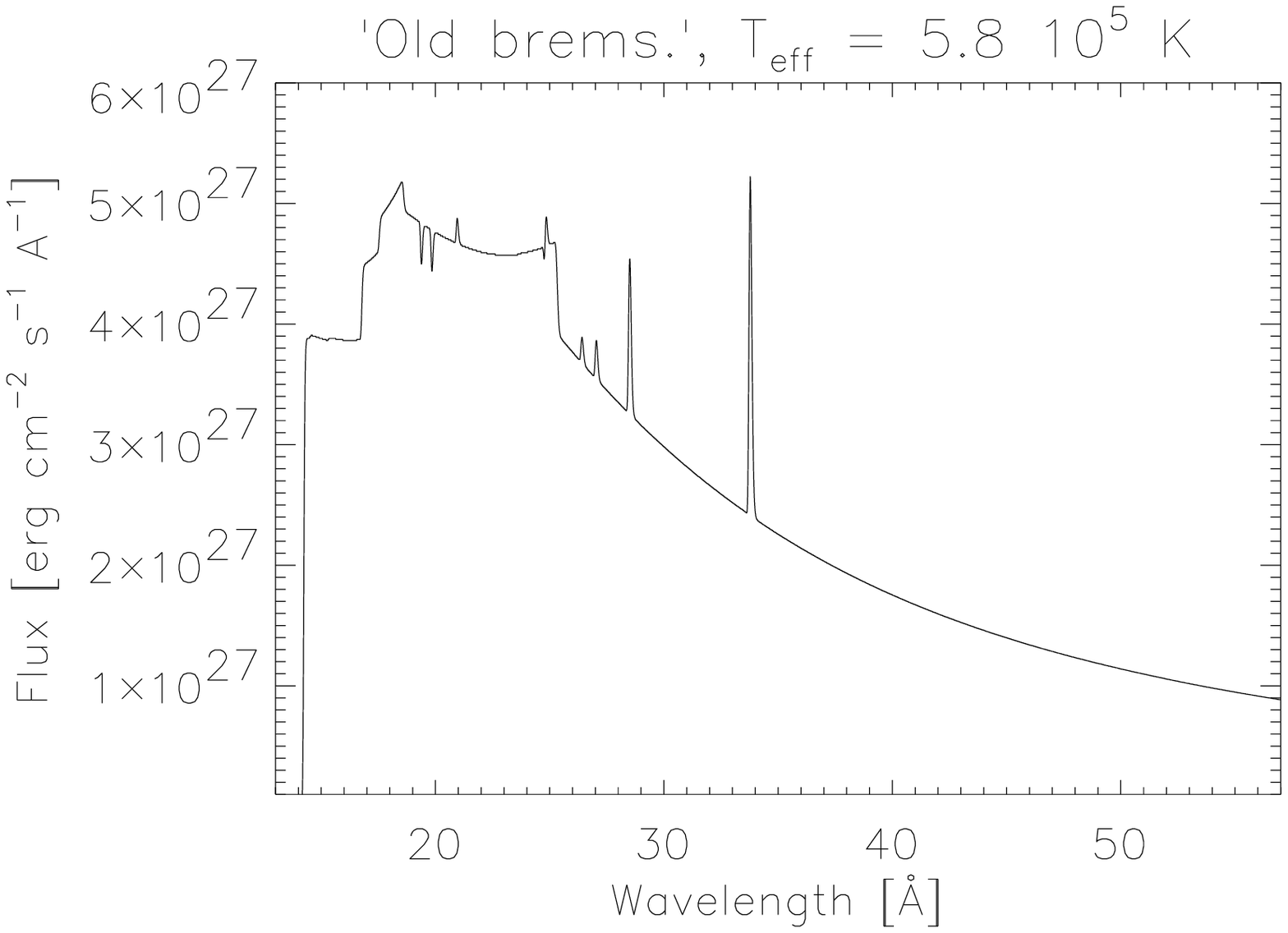}
  \includegraphics[width=8.4cm]{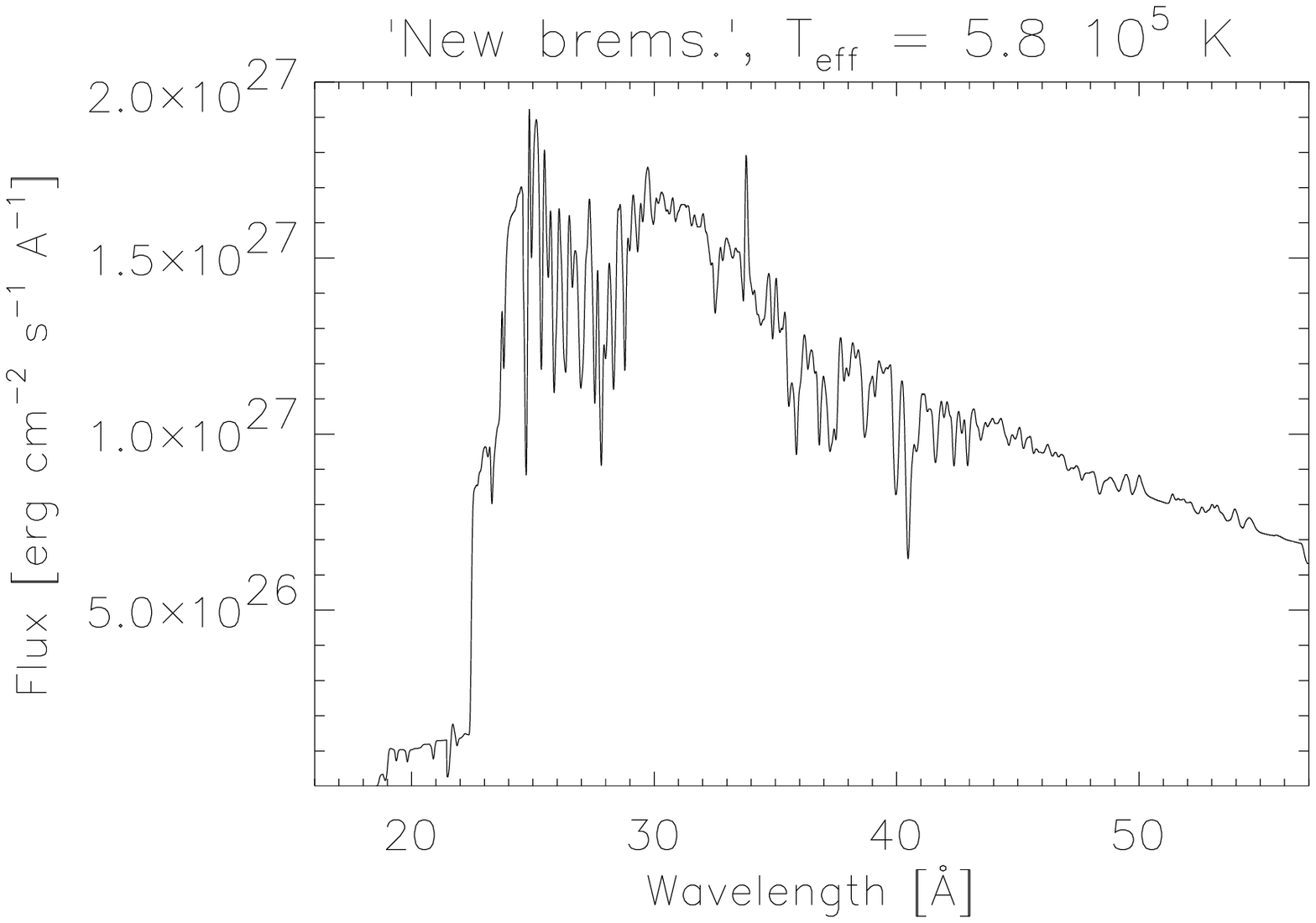}
  \caption[]{\label{models1} \emph{Upper}: Model spectrum calculated
  with the 'old bremsstrahlung' continuum. Because these are test
  models, there are only spectral lines from H, He, C, and N.
  \emph{Lower}: Model spectrum calculated with the 'new
  bremsstrahlung' continuum and spectral lines of H, He, C, N, O, and 
  Fe. There are mainly absorption lines in the spectrum.}
  \vfill
\end{figure}

\begin{figure}
  \includegraphics[width=8.4cm]{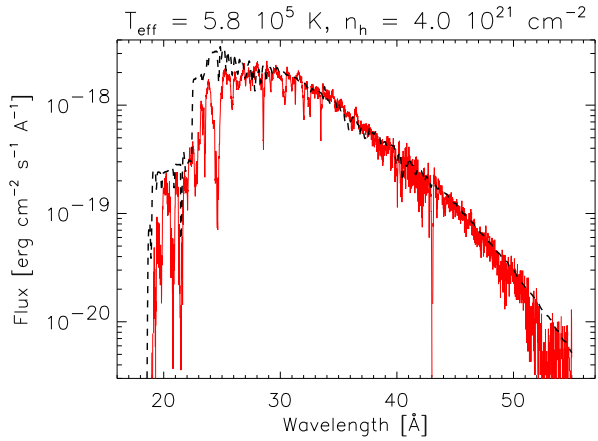}
  \caption[]{\label{models2} NLTE Model spectrum (calculated with the
  'new bremsstrahlung') with solar abundances plotted over the first
  15\,ks of the March observation. There is a very good agreement. A
  best--fit is provided with an effective temperature of
  $T_{\rm eff} = 5.8 \times 10^5$\,K
  ($L_{\rm bol} = 50,000 L_{\odot}$, $v_{\rm out} = 2500$\,km/s,
  $p_{\rm ref} = 10^{-2}$\,dyn/cm$^2$, $N = 3$). The absorption lines
  and the continuum absorption are too weak in the model. Interstellar
  absorption was taken into account
  ($n_h = 4.0 \times 10^{21}$\,cm$^{-2}$).}
\end{figure}

The {\tt PHOENIX} code package was modified to calculate model
atmospheres and synthetic spectra for the late phases of classical
novae where most of the energy is emitted in X--rays. Our models have
effective temperatures of $T_{\rm eff} = 10^5 - 10^6$\,K and a
bolometric luminosity of $L_{\rm bol} = 50,000 L_{\odot}$. In the
generation of the first test models the 'old bremsstrahlung' was
used. A typical NLTE model spectrum can be seen in the upper panel of
Fig.~\ref{models1}. There are only a few weak absorption lines and the
other spectral lines are in emission. Clearly, the observed spectra
(Fig.~\ref{observed_spectrum}) are not well represented by models
using the 'old bremsstrahlung'.\par
In calculating the other grids of models the 'new bremsstrahlung' was
used (Fig.~\ref{models1}, lower panel), where there are mostly
absorption lines in the spectrum. Considering the interstellar and
circumstellar absorption by hydrogen, the shape of the continuum model
spectrum is close to what is observed (Fig.~\ref{models2}). The
agreement between the continuum of the model and the observation is
acceptable, whereas there are differences in the spectral lines.\par
The best fit to the spectrum of nova V4743 is found at a temperature
of $T_{\rm eff} = 5.8 \times 10^5$\,K. To get the correct slope for
the pseudo--continuum (continuum formed by numerous overlapping
lines), a suitable value of the hydrogen column density has to be
used. A value of $n_h = 4.0 \times 10^{21}$\,cm$^{-2}$ for the
hydrogen column density provides the best fit, in contrast to the
value of $n_h = 1.41 \times 10^{21}$\,cm$^{-2}$ given by
\cite{dickey90} for the ISM. This discrepancy can be explained by
assuming additional absorption from the circumstellar shell, not
included in the value from the literature.\par
Close inspection of the measured spectrum reveals that some spectral
lines are not reproduced well or are missing in the model
spectrum. This is because we have used only solar abundances in the
model and have not enhanced abundances of, for example the CNO
elements, as is generally observed in novae and predicted by
theory. In the best model spectrum with solar abundances, all
absorption lines are too weak and there is too much emission around
24\,{\AA} ($\approx 517$\,eV). Increasing the abundances should
increase the absorption. We will do this in a succeeding
publication. There is an oxygen edge around 24\,{\AA}. Increasing the
O--abundance should add more opacity and could provide a better
fit. The computation of models with abundances deviating from solar
needs much more computational work and has not yet been done. With
such models it will be possible to estimate the chemical abundances of
the nova ejecta.\par
The lack of lines in the computed spectrum may also be caused
by using only line data from ions of H, He, C, N, and O and from Fe
XXI--XXVI in the model calculations. If the concentration of the
partial pressures of an ion is sufficiently large, all available
transitions between all available levels of these ions are treated in
NLTE. This procedure selects the last three ionization stages of C, N,
and O to be included in the NLTE calculation and the others of these
elements are treated in LTE.\par

\begin{figure}
  \includegraphics[width=7.2cm]{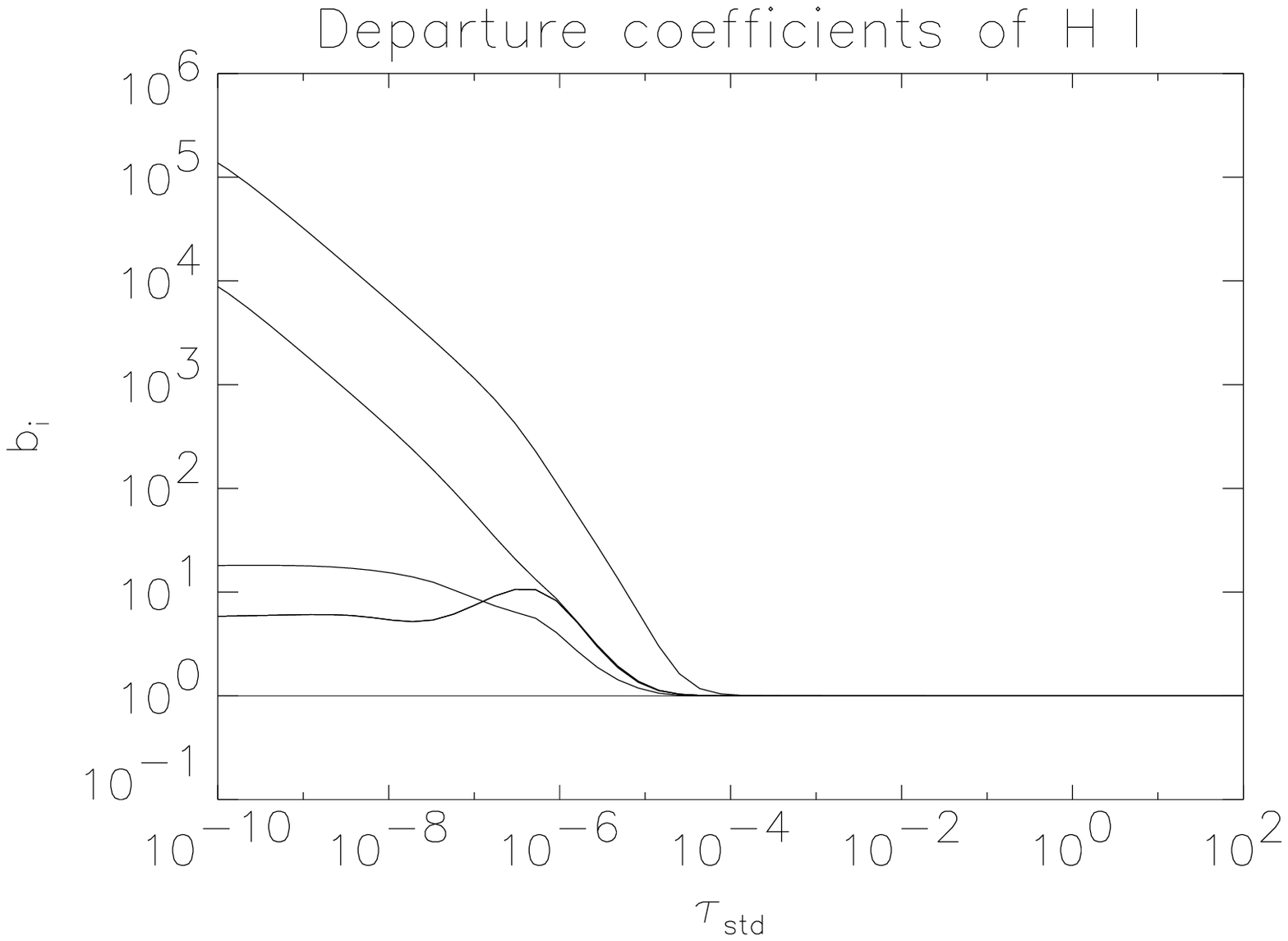}
  \includegraphics[width=7.2cm]{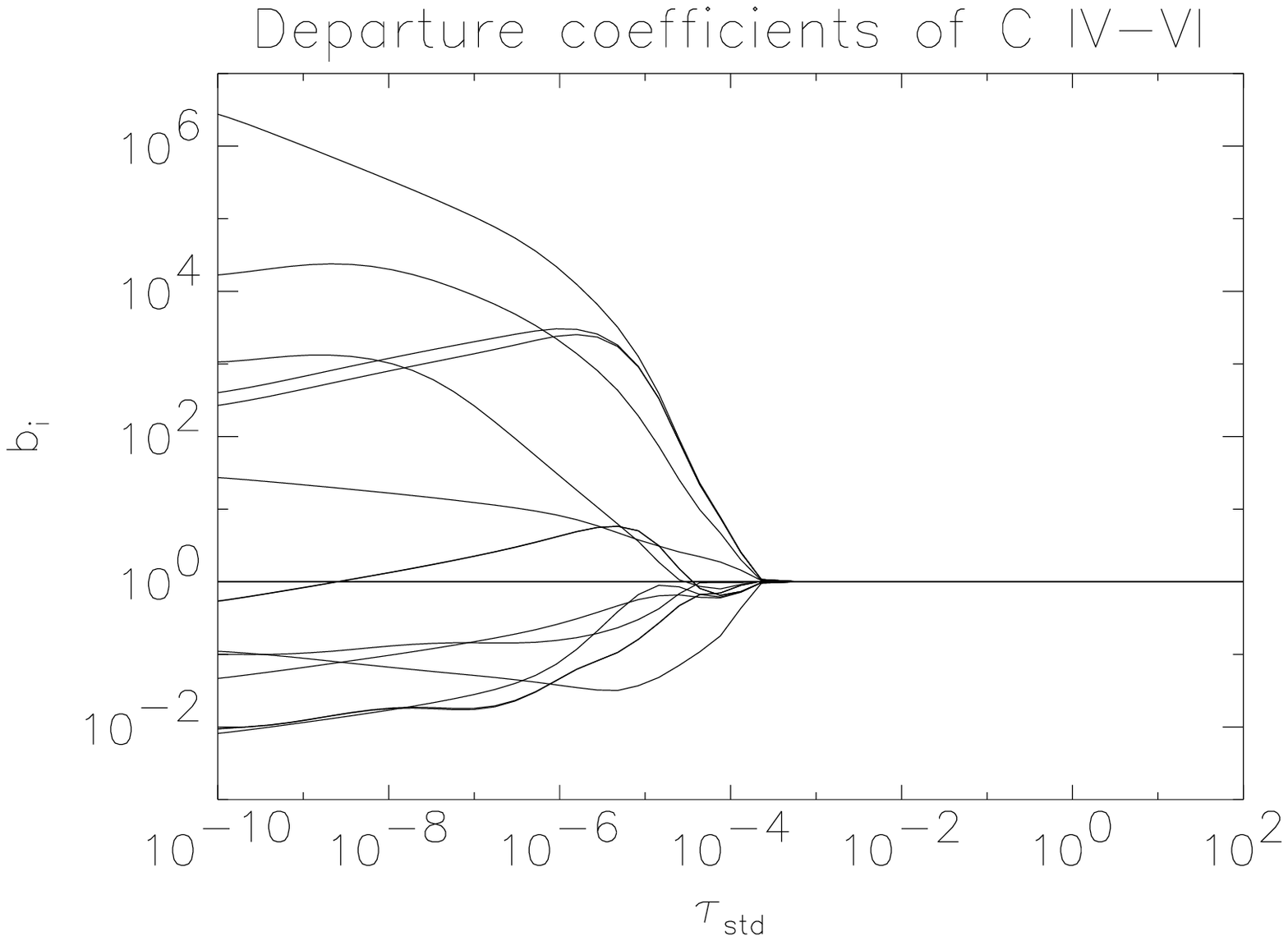}
  \includegraphics[width=7.2cm]{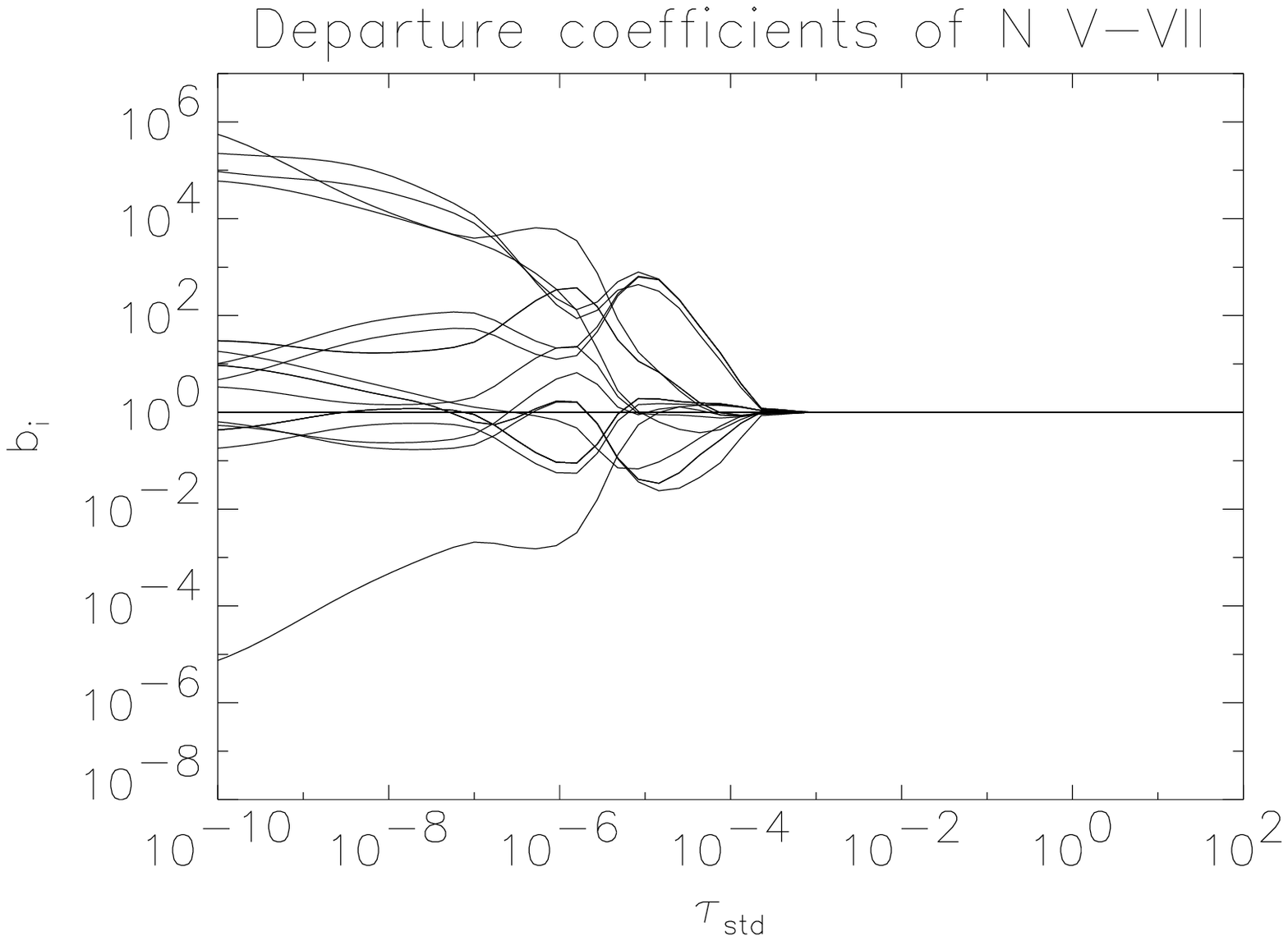}
  \includegraphics[width=7.2cm]{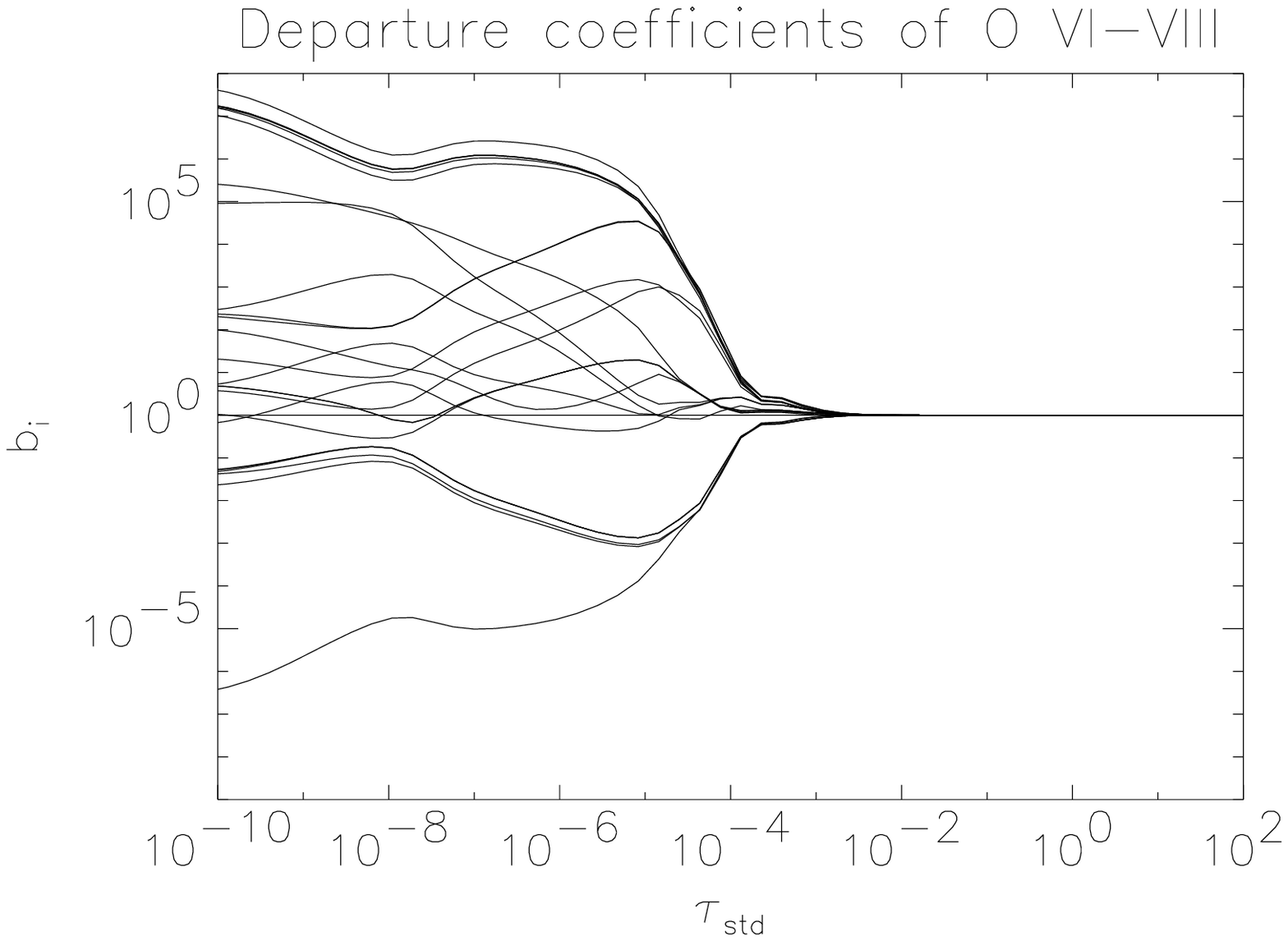}
  \caption[]{\label{depart_coeff} Departure coefficients $b_i$ from
  the best--fit model for the first 5 levels of the H-, C-, N-,
  and O--ions considered in these models. There are departure
  coefficients with
  values of 6 magnitudes larger or smaller than 1. For all elements
  the departures from LTE are strong. As expected, the departures are
  strongest for outer layers and there is no departure from LTE at
  large optical depths ($b_i = 1$).}
\end{figure}

\begin{figure}
  \includegraphics[width=8cm]{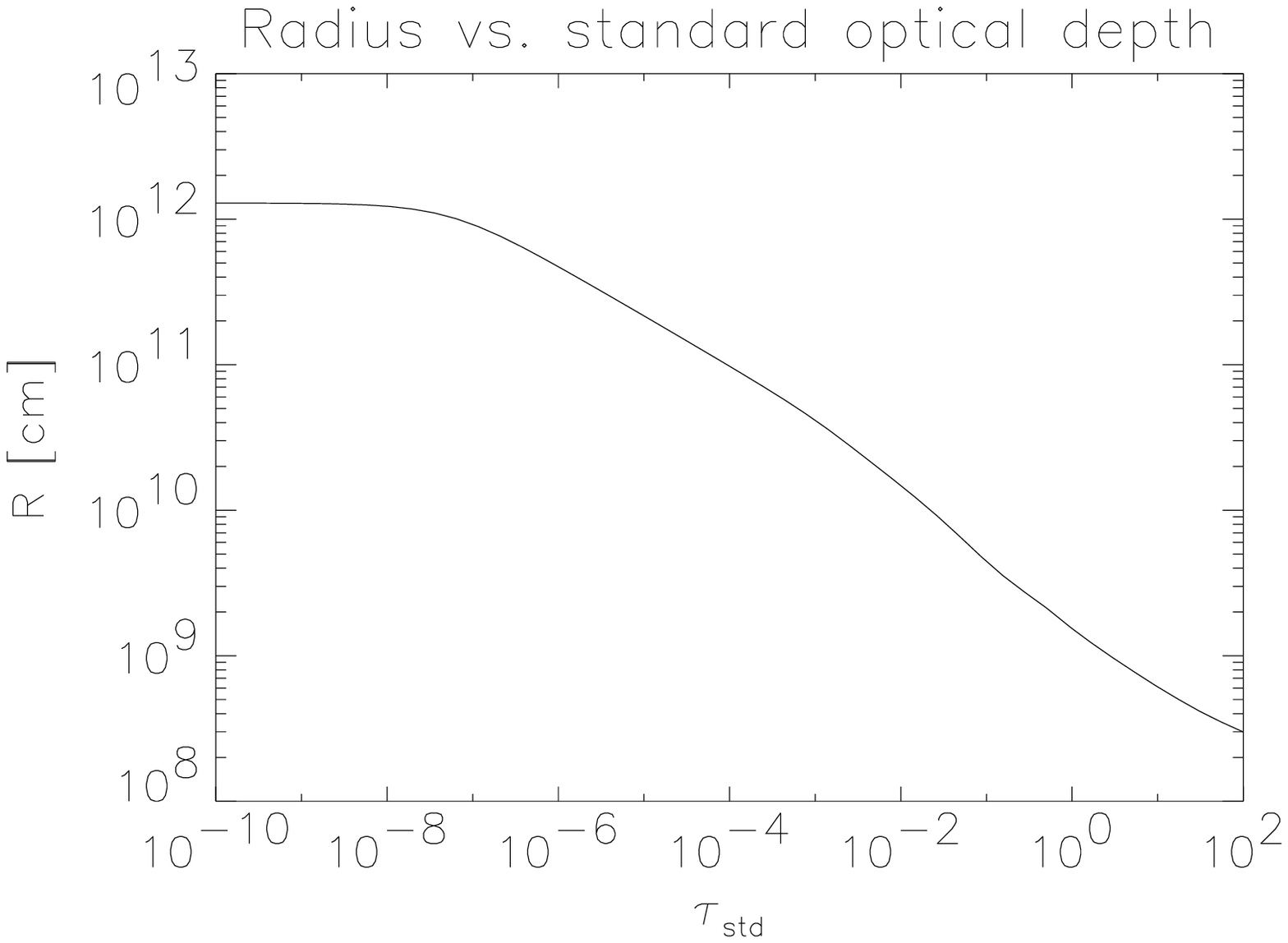}
  \includegraphics[width=8cm]{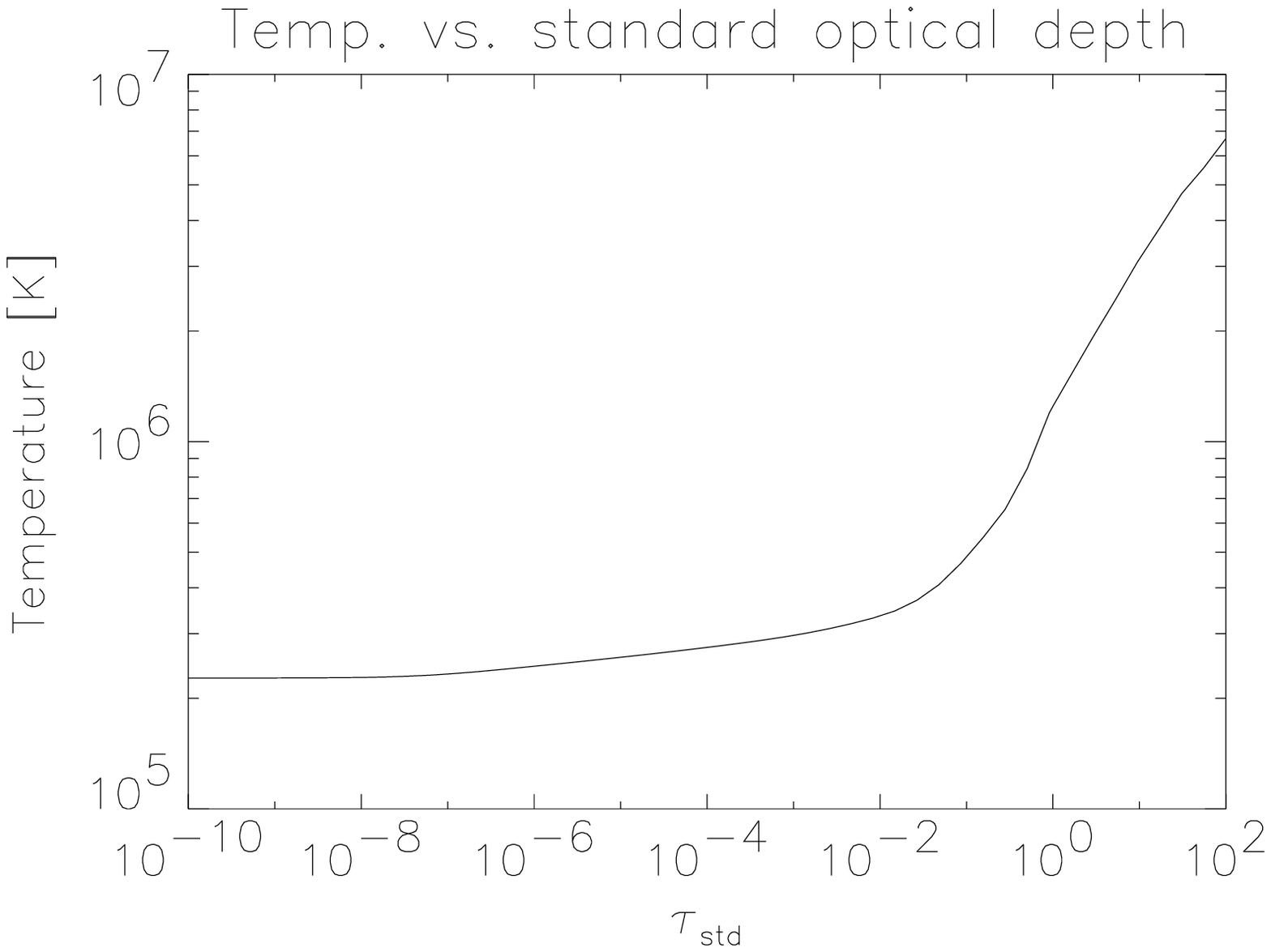}
  \caption[]{\label{tr_vs_tau} \emph{Upper}: Radius of the nova
  atmosphere plotted against the standard optical depth. The nova
  atmosphere is very extended, with a radius of
  $r \approx 10^{12}$\,cm at $\tau_{\rm std} = 10^{-12}$.\par
  \emph{Lower}: Temperature distribution in the atmosphere plotted
  against the standard optical depth. There is a wide variety of
  temperatures in the atmosphere. In the inner layers the temperature
  is several $10^6$\,K high.}
\end{figure}

\begin{figure}
  \includegraphics[width=8cm]{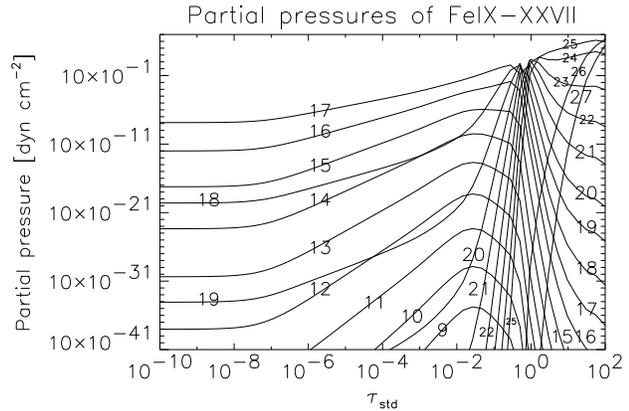}
  \caption[]{\label{partpr_Fe} Partial pressures of FeIX--XXVII. The
  highest ionization stages of iron exist at the deepest layers of the
  nova atmosphere. In the outer layers FeXVII is most abundant. The
  partial pressures of FeI to FeVIII are very small and lie below the
  range of the y--axis. Only ions from FeXV are abundant enough
  in the line forming region and therefore important.}
\end{figure}

For each ion it can be decided which atomic database will be
used. Because many X--ray spectral lines are required, the CHIANTI4 or
APED databases are used (Table \ref{xray_lines}). From a rough
comparison, the spectra calculated with each database are only
slightly different. The shape of the continuum and the number of
strong lines are the same. The quality of the databases has to be
determined for an accurate abundance analysis. For the models
presented in this paper we have used the APED database.\par
The main results from the models are that the X--ray emission is
dominated by thermal bremsstrahlung from the atmosphere surrounding
the WD, and that the hard spectral range of
$\lambda \lesssim 29$\,{\AA} (E $\gtrsim 428$\,eV) is dominated by
iron and nitrogen absorption. As expected, the departures from LTE are
large in the nova ejecta. To examine the strength of NLTE effects, the
departure coefficients can be plotted against the standard optical
depth $\tau_{std}$. The departure coefficient $b_i$ is the ratio of
the occupation numbers in level $i$ calculated in NLTE to calculations
in LTE for a specific ion. The stronger the departure from LTE the
more different the value of the departure coefficient from 1. In the
outer layers of nova atmospheres the departure coefficients can differ
up to six dex from 1 in both directions. This can be seen in
Fig.~\ref{depart_coeff} for different elements. Spectral lines are
formed in regions where the departure coefficients are very different
from 1. Therefore it is very important to treat all atomic transitions
in full NLTE.\par
In addition to strong departures from LTE, there are extreme physical
conditions in the nova ejecta. The atmosphere is very extended with a
radius of $r \approx 10^{12}$\,cm at $\tau_{\rm std} = 10^{-12}$
(Fig.~\ref{tr_vs_tau}, upper panel). There is a wide range of electron
temperatures (up to several $10^6$\,K, Fig.~\ref{tr_vs_tau}, lower
panel) and hence several of the highest ionization stages of an
element are simultaneously present in the ejecta. For example the
partial pressures of iron ions (Fe IX--XXVII) are plotted against the
standard optical depth in Fig.~\ref{partpr_Fe}. A similar effect has
already been observed in the IUE analysis of nova spectra
\cite[]{hauschildt92b}.\par

\section{Conclusion and outlook \label{concl}}

We have modified the {\tt PHOENIX} code to calculate model atmospheres
and synthetic spectra for the late X--ray phase of classical
novae. The agreement between the model and the observed spectrum of
nova V4743 Sgr is good in the continuum for an effective temperature
of $T_{\rm eff} = 5.8 \times 10^5$\,K if the 'new bremsstrahlung' (see
section \ref{addphys}) is used which is physically more reasonable. So
far, we only have computed model atmospheres with solar abundances for
computing time reasons and the fit to the spectral lines needs
to be improved by models with non--solar abundances. The hydrogen
column density of $n_h = 4.0 \times 10^{21}$\,cm$^{-2}$ providing the
best fit is high compared to $n_h = 1.41 \times 10^{21}$\,cm$^{-2}$
from \cite{dickey90}, because of additional absorption by the
circumstellar shell.\par
X--ray continuum emission is dominated by thermal bremsstrahlung and
its inverse processes, while bound--free emission is much weaker in
the nova atmosphere. In the hard spectral range of $\lambda \lesssim
29$\,{\AA} (E $\gtrsim 428$\,eV) iron and nitrogen absorption are
dominant. As expected, nova ejecta show strong departures from
LTE. They are very extended and several of the highest ionization
stages of all elements are simultaneously present.\par
In future work additional physical processes absorbing and emitting
X--rays will be implemented in the code. For example there are data
for dielectronic recombination and autoionization, and there are
better data for free--bound coefficients in the APED and CHIANTI4
databases.\par
Models with non--solar abundances will be calculated to obtain a
better fit for the line strengths and the continuum absorption and to
analyze the abundances of the nova ejecta. To determine reliable
chemical abundances more elements have to be included in the NLTE
calculations.\par
The data quality of the atomic databases as well as differences
between the CHIANTI4 and APED database will be checked, e.~g. by
examining the line profiles. By comparing model and observation, it
has to be determined which data of the individual ions reproduce the
spectral lines as accurately as possible. This is necessary for an
accurate abundance analysis.\par
The evolution of the X--rays should be examined, by modeling the
spectra of other observations of novae with the CHANDRA satellite. It
is important to test if it is possible to model the spectrum of novae
in the emission line phase, when nuclear burning has switched off and
the expanding shell is optically thin, like the spectrum of V382 Vel
\cite[]{burwitz02}. For this work more observations with CHANDRA are
necessary.\par

\begin{acknowledgements}
Some of the calculations presented here were performed at the
H\"ochstleistungs Rechenzentrum Nord (HLRN) and at the National Energy
Research Supercomputer Center (NERSC), supported by the U.S. DOE. We
thank all these institutions for a generous allocation of computer
time. Part of this work was supported by the DFG
(\it Deutsche Forschungsgemeinschaft\rm), project number
HA 3457/2--1. S. Starrfield was partially supported by grants from
NASA--CHANDRA, NASA--Theory, and NSF to ASU.
\end{acknowledgements}

\bibliography{xrcne}

\end{document}